\documentclass[11pt]{amsart}

\usepackage[usenames]{color}

\usepackage{amsmath}
\usepackage{amsthm}
\usepackage{amsfonts}
\usepackage{amssymb}
\usepackage{latexsym}
\usepackage{color}
\usepackage{graphicx}
\usepackage{float}

\usepackage{blindtext}

\usepackage{hyperref}
\usepackage[top=1in, bottom=1.5in, left=1in, right=1in]{geometry}

\newtheorem{problem}{Problem}
\newtheorem{example}{Example}

\newtheorem{definition}{Definition}

\newtheorem{remark}{Remark}

\newcommand{\Z}{\mathbb{Z}}

\newcommand{\F}{\mathbb{F}}

\begin{document}

\author[]{Nael Rahman}
\address{Department of Mathematics, The City College of New York, New York,
NY 10031} \email{naelrahman@gmail.com}

\author[]{Vladimir Shpilrain}
\address{Department of Mathematics, The City College of New York, New York,
NY 10031} \email{shpilrain@yahoo.com}

\title{MOBS: \lowercase{\uppercase{M}atrices \uppercase{O}ver \uppercase{B}it \uppercase{S}trings  public key exchange}}

%

\begin{abstract}
We use matrices over bit strings as platforms for Diffie-Hellman-like public key exchange protocols. When multiplying matrices like that, we use Boolean OR operation on bit strings in place of addition and Boolean AND operation in place of multiplication. As a result, (1) computations with these matrices are very efficient; (2) standard methods of attacking Diffie-Hellman-like protocols are not applicable.

\end{abstract}

\maketitle

\section{Introduction}

We consider $n \times n$ matrices whose entries are bit strings of the same fixed length $k$. When one multiplies  matrices like that, Boolean OR operation on bit strings is used in place of addition and Boolean AND operation is used in place of multiplication, as in the following

\begin{example}\label{ex1} Let $M= \left(\begin{array}{cc} (1 1 0) & (1 0 1) \\ (0 0 1) & (1 0 0)  \end{array}\right).$
Then $M^2= \left(\begin{array}{cc} (1 1 1) & (1 0 0) \\ (0 0 0) & (1 0 1)  \end{array}\right).$

\end{example}

With fixed $n$ and $k$, the set of all such matrices is a semigroup under multiplication. This semigroup is actually a monoid since it has the unit element. Denote the bit string of all 1s by $E$, and the bit string of all 0s by $O$. Then the matrix that has $E$ on the diagonal and $O$ elsewhere plays the role of the identity matrix; we denote it by $I=I_{n,k}$. The matrix $Z=Z_{n,k}$ that has $O$s everywhere plays the role of 0.

In a special case where $k=1$, we get matrices whose entries are single bits. We note that the algebra of these matrices is different from the algebra of matrices over the field $\F_2$ because in $\F_2$, one has 1+1=0 whereas with the Boolean operations, 1 OR 1 is 1. Furthermore, while $n \times n$ matrices  over $\F_2$ form a ring, our matrices form a {\it semiring} but not a ring since there are no additive inverses, e.g. there is no element $x$ such that $1+x=0$. We also note that not every (0-1) matrix that has a multiplicative inverse over $\F_2$ has a multiplicative inverse in our semiring. An example of such matrix would be $\left(\begin{array}{cc} 1 & 0 \\ 1 & 1 \end{array}\right).$ A system of linear equations corresponding to this matrix may have no solutions, or a unique solution, or more than one solutions, see Section \ref{telescoping}.  We therefore believe that Diffie-Hellman-like  protocols based on matrices over bit strings should not be susceptible to linear algebra attacks like e.g. \cite{Koblitz}.

In Section \ref{DH}, we describe a protocol, mimicking the standard  Diffie-Hellman public key exchange protocol \cite{DH}, using the semigroup of matrices over bit strings as the platform. Then we discuss the dynamics of the matrix entries generated by raising a given matrix to increasing powers and point out that this dynamics tends to be very simple in the sense that orbits generated by powers of a matrix tend to be rather short.

We therefore consider the protocol in Section \ref{DH} just a warm-up, and for the ``real thing" we use a semidirect product of the above mentioned semigroup with the group $S_k$ of permutations of the tuple $(1, \ldots, k)$. This is described in Sections \ref{Action} and  \ref{Protocol}.

Finally, we make one remark that is irrelevant to cryptography but may be of interest mathematically. Considering matrices over bit strings of length $k$, with Boolean operations as above, is equivalent to considering matrices over subsets of a set of $k$ distinct elements, with the union and intersection operations in place of the usual addition and multiplication, respectively. This is therefore a special case of a semiring of matrices over lattices; other special cases were considered in \cite{tropical} and \cite{semidirect tropical}.

\section{Diffie-Hellman-like protocol} \label{DH}

This is just  a straightforward mimicking of the standard  Diffie-Hellman public key exchange protocol \cite{DH}.

\begin{enumerate}

\item Alice and Bob agree on a $n \times n$ matrix $M$ whose entries are bit strings of the same fixed length
$k$. We will write the semigroup generated by $M$ multiplicatively.

\item Alice picks a random natural number $a$ and sends $M^a$ to Bob.

\item    Bob picks a random natural number $b$ and sends $M^b$ to Alice.

\item   Alice computes $K_A=(M^b)^a=M^{ba}$.

\item  Bob computes $K_B=(M^a)^b=M^{ab}$.
\end{enumerate}

Since $ab=ba$, both Alice and Bob are now in possession of the same
group element $K=K_A= K_B$ which can serve as the shared secret key.

\subsection{Dynamics} \label{Dynamics} Although standard attacks on Diffie-Hellman-like protocols are not applicable in our situation, there are other potential dangers. One thing to note is that different ``coordinates" in a bit string are independent under Boolean operations, i.e., our semigroup of matrices is a direct sum of $k$ semigroups of matrices whose entries are just single bits. When a matrix like that is raised to increasing powers, the number of different matrices obtained along the way may be rather small, which is not good for security. This is why we do not recommend using the above protocol in real life, but consider it just a warm-up for a more promising protocol in Section \ref{Protocol}.

That said, we would like to point out that it is not easy to quantify what is said in the previous paragraph.
Namely, given a matrix $M$ whose entries are just single bits, it is hard to tell how many different matrices there are in the set $\{M^s, s \in \Z_+\}$. Also of interest is the following

\begin{problem}
For a given $n \ge 2$, what $n \times n$ matrix $M$ whose entries are just single bits, gives the largest number of different matrices in the set $\{M^s, s \in \Z_+\}$? What is this largest number, as a function of $n$?
\end{problem}

Obviously, the largest number cannot be greater than $2^{n^2}$, the total number of different matrices with bit entries, but other than that, there are no general facts in this direction, to the best of our knowledge. From computer experiments, it looks like most orbits are very short, even for matrices of a large size, and typically, for a given matrix $M$ one of the following occurs very quickly:
\medskip

\noindent {\bf 1.} Powers of $M$ stabilize with some $M^s$ for a small $s$. This is the most popular case, and quite often, $M^2$ already has all entries equal to 1. To avoid this, one has to have an abundance of 0s in $M$.
Let $U$ denote the matrix of all 1s.
We have determined experimentally that for $3 \times 3$ matrices, the largest $s$ such that $M^s \ne U$ is $s=4$.
One of the corresponding matrices $M$ is $\left(\begin{array}{ccc} 0 & 1 & 0\\ 1 & 0 & 1\\ 1 & 0 & 0 \end{array}\right)$. There are six $3 \times 3$ matrices like that altogether.

With $4 \times 4$ matrices, the largest $s$ such that $M^s \ne U$ is $s=9$. There are 24 matrices like that; one of them is $\left(\begin{array}{cccc} 0 & 1 & 1 &0\\ 0 &0 & 1 &  0\\ 0 & 0 & 0& 1\\ 1 & 0 & 0 & 0\end{array}\right)$.

With $5 \times 5$ matrices, the largest $s$ such that $M^s \ne U$ is $s=16$. There are 120 matrices like that; one of them is $\left(\begin{array}{ccccc}  0 & 1 & 1 &0 &0\\ 0 &0 & 1 &  0 &0\\ 0 & 0 & 0& 1& 0\\ 0 & 0 & 0 & 0 &1\\ 1 & 0 & 0 & 0 & 0 \end{array}\right)$.

Based on these small values of $n$, one can make a naive conjecture on $s=s(n)$ being very close to $2^{n-1}$ for $n  \times n$ matrices.
\medskip

\noindent {\bf 2.} Powers of $M$ do not stabilize, but ``oscillate", i.e., for some small $r$ and $s$, one has $M^{r+s}=M^r$, i.e., there are only short orbits. An example would be $M = \left(\begin{array}{ccccc} 0 & 0 & 0 & 1& 1 \\0 & 0 & 0 & 0& 1 \\0 & 1 & 1 & 1& 0 \\ 1& 0 & 0 & 0& 1\\ 0 & 1 & 0 & 0 & 0\end{array}\right).$ For powers of this $M$, we have $M^{5+2k} = M^{5}$ for any $k \ge 0$, but $M^{2m} \ne M^5$ for any $m \ge 0$.



\section{Semidirect products}
\label{Semidirect}

We now recall the definition of a semidirect product:

\begin{definition} Let $G, S$ be two groups,  and let the group $S$ act on $G$ by automorphisms.
Then the
semidirect product of $G$ and $S$ is the set
$$\Gamma = G \rtimes S = \left \{ (g, h): g \in G, ~h \in S \right \}$$
with the group operation given by\\
\centerline{$(g, h)(g', h')=(h'(g) \cdot  g', ~h \cdot h')$.}\\
Here $h'(g)$ denotes the result of the action on $g$ by
$h'$, and when we write a product $h \cdot h'$ of two
morphisms, this means that $h$ is applied first.
\end{definition}

One can also use this construction if $G$ is not necessarily a
group, but just a semigroup, and/or consider endomorphisms (i.e.,
self-homomorphisms) of $G$, not necessarily automorphisms. Then the
resulting  semidirect product will be a semigroup, not a group, but this is sufficient for being the  platform of a Diffie-Hellman-like key exchange protocol.

Semidirect products of (semi)groups have been previously used as platforms for public key establishment protocols, see e.g. \cite{Eraser}, \cite{semidirect tropical}, \cite{Habeeb}, \cite{semidirect survey}, \cite{MAKE}. The platform we use in this paper is novel; it is described in the next Section \ref{Action}.

\section{Our platform} \label{Action}

Our platform semigroup will be a semidirect product of two semigroups. One of them, call it $G$, is the semigroup (under multiplication operation) of matrices  over bit strings of the same fixed length $k$. The other one, call it $S$, is the semigroup (it is actually a group) of permutations of $k$ distinct objects.

The action of $S$ on $G$ will be as follows. A permutation $h \in S$ acts on each bit string in a matrix
$M \in G$ by permuting its $k$ bits. In reference to our Example \ref{ex1}, here is how this works.

\begin{example}\label{ex2} Let $M= \left(\begin{array}{cc} (1 1 0) & (1 0 1) \\ (0 0 1) & (1 0 0)  \end{array}\right).$ Let the permutation $h$ of 3 objects be $(a ~b ~c) \to (c ~a ~b)$. Then
$h(M)= \left(\begin{array}{cc} (0 1 1) & (1 1 0) \\ (1 0 0) & (0 1 0)  \end{array}\right),$ and

\centerline{$(M, h)^2 = (M, h)(M, h)=(h(M) \cdot  M, ~h^2) = (\left(\begin{array}{cc} (0 1 0) & (1 0 1) \\ (1 0 0) & (1 0 0)  \end{array}\right), ~h^2)$}

\end{example}

\section{Protocol description}\label{Protocol}

Below is the protocol description. Parameters are discussed separately, in Section \ref{Parameters}.
\smallskip

\begin{itemize}

\item[1.] ({\it key selection}) {\bf (i)} Alice and Bob agree on a matrix $M$ over bit strings of the same fixed length $k$ and on a  permutations $h$ of the tuple $(1, \ldots, k)$.

\smallskip

\noindent {\bf (ii)} Alice selects a private integer $a$ and Bob selects a private integer $b$.

\smallskip

\item[2.]   Alice computes $(M, h)^a$ and sends {\bf only the first component} (call it $A$) of the
result to Bob.

\smallskip

\item[3.]   Bob computes $(M, h)^b$ and sends {\bf only the first component} (call it $B$) of the
result to Alice.

\smallskip

\item[4.]  Alice computes $(B, x) \cdot (A, ~h^a) = (h^a(B)\cdot A, ?)$. Her key is now $K_A = h^a(B)\cdot A$.
\smallskip

\item[5.]   Bob computes $(A, y) \cdot (B, ~h^b) = (h^b(A)\cdot B, ?)$.  His key is now $K_B =h^b(A)\cdot B$.

\smallskip

\item[6.]  Since $(M, h)^{a+b} = (B, x) \cdot (A, ~h^a) = (A, ~y) \cdot (B, ~h^b) =
(K, ~h^{a+b})$, we should have $K_A = K_B = K$, the shared secret
key.

\end{itemize}

\begin{remark}
Note that, in contrast with the ``standard" Diffie-Hellman key
exchange, correctness here is based on the equality $x^{m}\cdot
x^{n} = x^{n} \cdot x^{m} =  x^{m+n}$  rather  than on the equality
$(x^{m})^{n} = (x^{n})^{m} = x^{mn}$. In  the ``standard"
Diffie-Hellman set up, our trick would not work because, if the
shared key $K$ was just the product of two  openly transmitted
elements, then anybody, including the eavesdropper, could compute
$K$.
\end{remark}

\section{Parameters and keys sampling}
\label{Parameters}

Suggested parameters are: $n=3$, $k=381$ (the reason for selecting this $k$ is explained below). With these parameters, the size of the public matrix $M$ is 3,429 bits. Bits in the matrix $M$ are selected so that each time, the ``1" bit is selected with probability $p$ and the ``0" bit is selected with probability $1-p$. We suggest $p=\frac{1}{2}$, although other values are possible, see discussion in Section \ref{Security}.

The suggested size of private exponents $a$ and $b$ is 500 bits. Sampling of either key is done by selecting a binary number, uniformly at random, from the set of all 500-bit numbers.

One of the important points is that the order of the (public) permutation $h$ should be rather large; otherwise the orbits under the action by this permutation will be too short.

The largest order of an element of the group  $S_k$ of permutations on $k$ distinct objects is known as Landau's function $g(k)$ \cite{Landau}. It is known that this function grows faster than $e^{\sqrt{k}}$.  To actually build a permutation of a large order, one can do the following. The permutation $h$ will be a product of cycles of different prime lengths. Thus, start by computing the sum of primes $2+3+5+7+11+\ldots$  up to some point. For example, the sum of primes up to 53 is equal to $k=381$. The product of these primes is about $3.26 \cdot 10^{19} \approx 2^{65}$, which is large enough.

Now we build the permutation $h$ on $\{1, \ldots, k\}$, going ``left to right", as a product of cycles $(1 ~2), (3 ~4 ~5), (6 ~7 ~8 ~9 ~10), \ldots$ of increasing prime lengths, until we exhaust all $k$ integers.

\section{Security}
\label{Security}

First we note that transmitted matrices (at Steps 2 and 3 of the protocol) are:
\begin{align} \label{A}
A = h^{a-1}(M) \cdot h^{a-2}(M) \cdots h(M) \cdot M.
\end{align}

\begin{align} \label{B}
B = h^{b-1}(M) \cdot h^{b-2}(M) \cdots h(M) \cdot M.
\end{align}

\noindent Our security assumption is that it is computationally infeasible to recover the shared key

$$K = h^{a+b-1}(M) \cdot h^{a+b-2}(M) \cdots h(M) \cdot M$$

\noindent from $M, A,$ and $B$.

We do not claim that the shared key $K$ is a matrix that is indistinguishable from a random matrix over bit strings of length $k$. In fact, if bits in the entries of the public matrix $M$ are selected so that the ``1" bit is selected with probability $\frac{1}{2}$, then, according to our computer simulations, in the matrix $K$ the ``0" bits prevail. On average, about 67\% of bits in $K$ are 0s. There is still a huge number of matrices like that, so if $K$ is indistinguishable from other matrices with
67\% of 0 bits, this is still fine as far as security is concerned. 

Alternatively, when selecting bits in the matrix $M$, one can select ``1" bits with probability $p>\frac{1}{2}$; then, with appropriate $p$, the matrix $K$ will have (on average) an equal number of ``1" and ``0" bits. According to our computer simulations, this ``appropriate $p$" is about  0.535. Figure 1 below shows a scatter plot of the ratio of ``0" bits in the matrix $K$ depending on $p$.

\begin{figure}[h!]\label{plot}
\vskip -0.2cm
\includegraphics[width=3in, height=3in]{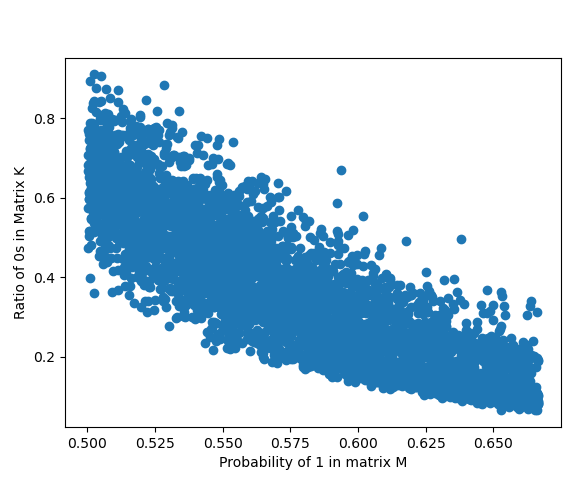}
\vskip -0.5cm \caption{Ratio of 0s in the matrix $K$} \label{bins}
\end{figure}

\subsection{The ``telescoping" attack and linear equations} \label{telescoping}

The following ``telescoping" attack is similar to the attack on the MAKE protocol \cite{MAKE} offered in \cite{Koblitz}. From
(\ref{A}), we get after applying $h$ to both sides:

\begin{align} \label{hA}
h(A) \cdot M = h^a(M) \cdot A.
\end{align}

All matrices in (\ref{hA}) are public, except the matrix $h^a(M)$. To use this equality to break our scheme, one has to address the following two problems:

\medskip

\noindent {\bf 1.} Solve the linear equation (\ref{hA}) for $h^a(M)$. One can also consider this a {\it system} of linear equations in the entries of $h^a(M)$.
\medskip

\noindent {\bf 2.} Recover $h^a$ from $h^a(M)$ and $M$. This is sufficient to break the scheme since $K=h^a(B)\cdot A$.
\medskip

\noindent  The first problem can be solved by ``brute force" because, as we have mentioned in Section \ref{Dynamics},
our semigroup of matrices is a direct sum of $k$ semigroups of matrices whose entries are just single bits. Thus, the problem reduces to a linear matrix equation where matrix entries are just single bits. Therefore, if matrices are of a small size, one can just try all possible matrices to find a solution. The problem, however, is that a solution may not be unique, and even if there are just 2 solutions for matrices over single bits, then for matrices over bit strings of length $k$ there will be $2^k$ different solutions. For the value of $k=200$ that we suggest, it will therefore be infeasible to find all solution of the form $\sigma(M)$ for some permutation $\sigma$, and without this knowledge one cannot address problem (2) above.

Here is a small example to illustrate how one gets non-unique solutions.

\begin{example}\label{ex solutions} Suppose we have the following matrix equation: $X \cdot \left(\begin{array}{cc} 1 & 0\\ 1 & 1  \end{array}\right) = \left(\begin{array}{cc} 1 & 1\\ 1 & 0  \end{array}\right)$. Then both  matrices below are solutions:
\medskip

\noindent $X_1 =  \left(\begin{array}{cc} 0 & 1\\ 1 & 0  \end{array}\right), ~X_2 =  \left(\begin{array}{cc} 1 & 1\\ 1 & 0  \end{array}\right).$

\end{example}

\noindent On the other hand, the following matrix equation has no solutions:
\medskip

\noindent $X \cdot \left(\begin{array}{cc} 1 & 0\\ 1 & 1  \end{array}\right) = \left(\begin{array}{cc} 0 & 1\\ 1 & 1  \end{array}\right)$.
\medskip

\noindent Finally, the following matrix equation has a unique solution: 
\medskip

\noindent $X \cdot \left(\begin{array}{cc} 1 & 0\\ 1 & 1  \end{array}\right) = \left(\begin{array}{cc} 1 & 0\\ 0 & 0   \end{array}\right)$.

\noindent The unique solution is $X =  \left(\begin{array}{cc} 1 & 0\\ 0 & 0  \end{array}\right)$.

%
%
%
%
%

\section{Implementation and performance}
\label{Implementation}

The scheme of this paper was implemented using Python. The code is available online, along with a challenge, see \cite{Python2}.

This implementation is rather efficient, due to the fact that there are no ``actual" multiplications, just Boolean operations.
On a regular desktop computer with Intel Gemini Lake 2 GHz processor, without any optimization or parallelization, the runtime is about 10 sec, and it can be brought down to under 1 sec when optimized.

\section{Conclusions}

$\bullet$ We have offered a key exchange protocol, resembling the classical Diffie-Hellman protocol, based on a semidirect product of a cyclic semigroup of matrices over bit strings (with Boolean operations) and a cyclic group of permutations.

$\bullet$ Security assumption, analogous to the {\it computational Diffie-Hellman assumption}, is computational infeasibility of recovering a matrix (over bit strings), which is a product of $(a+b)$ matrices over permuted  bit strings, from two matrices, one of which is a product of $a$ matrices and the other a product of $b$ matrices over permuted bit strings.

\vskip .5cm

\noindent {\it Acknowledgement.} We are grateful to Ethan Akin and Dima Grigoriev for helpful comments.

\end{document}